\documentclass[epj]{svjour}
\usepackage{graphics}
\usepackage[latin1]{inputenc}

\usepackage{amsmath}
\usepackage{rotating}
\usepackage{psfrag}
\usepackage{color}
\usepackage{ifthen}

\pdfoutput=1
\usepackage{graphicx}
\usepackage{dcolumn}
\usepackage{bm}
\usepackage{amsmath}
\usepackage{amssymb}
\usepackage{dsfont}
\usepackage{amsfonts}
\usepackage[UKenglish]{babel}
\usepackage{hyperref}
\usepackage{nicefrac}
\usepackage{verbatim}
\usepackage{bbm}
\usepackage{color}
\usepackage{multirow}
\begin{document}
\title{Can we decide whether QCD is confining or not at high temperature?}

\author{L. Ya. Glozman  
}                     
\institute{Institute of Physics,  University of Graz, A--8010 Graz, Austria}
\date{Received: date / Revised version: date}
%
\abstract{At high temperature measurements of the Polyakov loop suggest a deconfinement
transition to the (strongly interacting) quark-gluon plasma.
At the same time at the infinitely large temperature the four-dimensional QCD
is reduced to the three-dimensional QCD that is confining. The Polyakov loop
and related $Z_3$ symmetry are strict order parameters only for infinitely
heavy quarks. In such a situation the  $SU(2)_{CS}$ and
$SU(4)$ symmetries of confinement in the light quark sector could be helpful to distinguish between
the confining and  deconfining phase in a regime where 
$SU(2)_L \times SU(2)_R$ and $U(1)_A$ symmetries are manifest. In order
to reveal a presence or absence of these symmetries one needs to measure
and compare correlation functions related by these symmetry transformations.
} 

\maketitle
\section{Introduction}

In spite of big efforts, both experimental and theoretical, 
to reveal properties of QCD at high temperature, the issue 
about microscopical structure of the matter is still open. Experimentally a transition
to the (strongly interacting) "quark-gluon plasma" is assumed, given
observation of some specific properties of the matter
at high temperature in heavy ion collisions. What is  reliably established
on the
lattice -
is a crossover to the chirally symmetric regime. 
It has recently been demonstrated by the JLQCD collaboration
that at $T > T_c$ not only $SU(2)_L \times SU(2)_R$ but also a $U(1)_A$ symmetry is restored \cite{JLQCD}.

Situation with confinement is by far not clear, however. On the
one hand, lattice simulations of the Polyakov loop \cite{Polyakov} suggest
a transition to the deconfinement regime approximately at the
same temperatures like chiral restoration, see for a review Ref. \cite{K}. On the other hand, the Polyakov
loop and related $Z_3$ symmetry can be considered as order
parameters for confinement only for infinitely heavy quarks. At the same time it is known
that at the infinitely high temperature QCD becomes effectively a three-dimensional
theory  which is known to be confining \cite{AP}.

In this short note we suggest that  $SU(2)_{CS}$ and $SU(4)$ symmetries
of  confinement in the light quark sector \cite{G}\footnote{These
symmetries have been discovered given a large degeneracy of hadrons 
observed in lattice simulations upon
elimination of the quasi-zero modes from the quark propagators\cite{DGL1,DGL2,DGP1,DGP2}.}
could serve as a confinement-deconfinement order parameter in a regime
where chiral $SU(2)_L \times SU(2)_R$ and $U(1)_A$ symmetries are
manifest. 
Then, through a study of the correlation functions
 that are connected by the $SU(2)_{CS}$ and $SU(4)$
transformations and not linked by the $SU(2)_L \times SU(2)_R$ and $U(1)_A$ symmetries one could judge about existence or nonexistence of the
$SU(2)_{CS}$ and $SU(4)$ symmetries at high temperature.

\section{$SU(2)_{CS}$ and $SU(4)$ symmetries of confinement }

Consider the QCD Hamiltonian in Coulomb gauge \cite{CL}:

\begin{equation}
H_{QCD} = H_E + H_B
\nonumber
\end{equation}

 \begin{equation}
 + \int d^3 x \Psi^\dag({\boldsymbol{x}}) 
[-i \boldsymbol{ \alpha} \cdot \boldsymbol{\nabla} + \beta m ]  \Psi(\boldsymbol{x})
+ H_T + H_C,
\end{equation}

\noindent
where the transverse (magnetic) and Coulombic interactions are:

\begin{equation}
H_T = -g \int d^3 x \, \Psi^\dag({\boldsymbol{x}}) \boldsymbol{\alpha} 
\cdot t^a \boldsymbol{A}^a(\boldsymbol{x}) \, \Psi(\boldsymbol{x}) \; , 
\end{equation}

\begin{equation} 
H_C = \frac{g^2}{2} \int  d^3 x \, d^3 y\, J^{-1} \ \rho^a(\boldsymbol{x})  F^{ab}(\boldsymbol{x},\boldsymbol{y}) \, J \, \rho^b(\bf y) \; ,
\end{equation}
 
\noindent
with $J$ being  Faddeev-Popov determinant, $\rho^a(\boldsymbol{x})$
is a color-charge density and $F^{ab}(\boldsymbol{x},\boldsymbol{y})$ is a confining
Coulombic kernel.

The fermionic and transverse parts of the Hamiltonian
have the $SU(2)_L \times SU(2)_R$ and $U(1)_A$ symmetries.
A symmetry of the confining Coulombic part is higher, however.
It is not only invariant under the $SU(2)_L \times SU(2)_R$ and $U(1)_A$ transformations, like the fermionic and magnetic parts, but is also a singlet
with respect to  $SU(2)_{CS}$ {\it chiralspin} rotations as well as 
$SU(4)$ transformations \cite{GP}. 

The chiralspin $SU(2)_{CS}$ transformations are defined as rotations
of the fundamental vectors

\begin{equation}
U = (u_L,   u_R)^T ~~~~~~~
D = (d_L,  d_R)^T 
\end{equation}

\noindent
in an imaginary chiralspin space:

\begin{equation}
 U \rightarrow  U' = e^{i \frac {\boldsymbol{\varepsilon} \cdot \boldsymbol{\Sigma}}{2}} U\; ,~~~~~~
 D \rightarrow  D' = e^{i \frac {\boldsymbol{\varepsilon} \cdot \boldsymbol{\Sigma}}{2}} D \; ,
 \end{equation}

\noindent
where $\boldmath{\Sigma}$ are $ 4 \times 4$ matrices  

\begin{equation}
\boldsymbol{\Sigma} = \{ \gamma^0, i \gamma^5 \gamma^0, -\gamma^5 \} \;, 
\end{equation}

\noindent
that satisfy the $SU(2)$ algebra:

\begin{equation}
[\Sigma^i,\Sigma^j] = 2 i \epsilon^{i j k} \, \Sigma^k . 
\end{equation}
 
\noindent
Upon rotations in the chiralspin space the left- and right-handed
components of the quark fields get mixed.

 A group that contains at the same time $SU(2)_L \times SU(2)_R $
and $SU(2)_{CS} \supset U(1)_A$ is $SU(4)$ with the fundamental vector

\begin{equation}
\Psi = (u_{\textsc{L}}, u_{\textsc{R}},   d_{\textsc{L}},  d_{\textsc{R}} )^T 
\end{equation}

\noindent
and a set of generators

\begin{equation}
 \{(\tau^a \otimes \mathds{1}_D), (\mathds{1}_F \otimes \Sigma^i), (\tau^a \otimes \Sigma^i) \} . 
 \end{equation}

\section{What symmetries can we expect at high temperatures? }

In the plasma (deconfining) regime one
expects a priori that the system has the $SU(2)_L \times SU(2)_R$ (or $SU(2) \times SU(2) \times U(1)_A$) symmetries of the QCD Lagrangian. One views the
deconfined  plasma as a system of quarks and gluons  that freely
propagate through the matter. What is generic for plasma is a Debye screening
of the electric charge. At large separation between the quasiparticles
the colour-Coulombic
interaction between the quasiparticles is  screened, but the magnetic interaction is expected to be  there. Consequently, if one deals with the
deconfined plasma, correlators should  reveal absence
 of the  $SU(2)_{CS}$ and $SU(4)$ symmetries. 
Even without the Debye screening, if we treat the system perturbatively,
a relevant symmetry cannot be higher than the $SU(2) \times SU(2) \times U(1)_A$ symmetry of  the QCD Lagrangian,
because   magnetic part of perturbative interactions manifestly breaks
the $SU(2)_{CS}$ and $SU(4)$ symmetries of  Coulombic interaction.

In contrast, if one deals with the system with confinement and if 
at the same time
the
$SU(2)_L \times SU(2)_R \times U(1)_A$ symmetry is restored, the correlators
should reveal the $SU(2)_{CS}$ and $SU(4)$ symmetries, like it happens
in hadrons at zero temperature upon the near-zero-mode elimination 
\cite{DGL1,DGL2,DGP1,DGP2}.

\section{What correlators should be measured and compared?}

In order to address the question of $SU(2)_{CS}$ and $SU(4)$
symmetries one needs to study correlators that are connected
by $SU(2)_{CS}$ and $SU(4)$ transformations and at the same
time do not transform into each other under 
$SU(2)_L \times SU(2)_R \times U(1)_A$. There are many possible 
suitable operators
\cite{GP,DGP2}, but for  practical reasons it is convenient to
choose channels without the disconnected contributions of valence quarks.
For example, the $J=1$ $\rho$-meson operators

\begin{equation}
\bar \Psi \tau^a \gamma^i \Psi, ~~ \bar \Psi \tau^a \gamma^0 \gamma^i \Psi 
\end{equation}
belong to the same triplet of the chiralspin group and are members of
a 15-plet of the $SU(4)$ group. Coincidence of their correlators is a sufficient
condition to claim $SU(2)_{CS}$ and $SU(4)$ symmetries.
At the same time, their unequility would imply absence
of $SU(2)_{CS}$ and $SU(4)$ symmetries.

In the baryon sector convenient operators could e.g. be

\begin{equation}
 \mathcal{O}_{N^{\pm}} = \varepsilon^{abc} \mathcal{P}_{\pm} u^a \left[ u^{b T} C \gamma_5 d^{c}  \right] 
\end{equation}

\noindent
and

\begin{equation}
 \mathcal{O}_{N^{\pm}} = i \varepsilon^{abc} \mathcal{P}_{\pm} u^a \left[ u^{b T} C \gamma_5 \gamma_0 d^{c}  \right],
 \end{equation} 
 that belong to distinct representations of $SU(2)_L \times SU(2)_R$ and $U(1)_A$
 groups and at the same time are members of the same irreducible representations of  $SU(2)_{CS }$ and of $SU(4)$.

\medskip
 Partial support from the
Austrian Science Fund (FWF) through  grant
P26627-N27 is acknowledged.

\end{document}